\documentclass[journal,twoside,web]{ieeecolor}
\usepackage{lcsys}
\usepackage{cite}
\usepackage{amsmath,amssymb,amsfonts}
\usepackage{algorithmic}
\usepackage{graphicx}
\usepackage{textcomp}
\def\BibTeX{{\rm B\kern-.05em{\sc i\kern-.025em b}\kern-.08em
    T\kern-.1667em\lower.7ex\hbox{E}\kern-.125emX}}
\let\labelindent\relax

\usepackage{cite}
\usepackage{amsmath,amssymb,amsfonts}
\usepackage{bm, hyperref}
\usepackage{nicefrac}
\usepackage[table]{xcolor}
\usepackage{enumitem}
\usepackage[caption=false, font=footnotesize]{subfig}
\usepackage{booktabs}

\definecolor{darkgreen}{rgb}{0.0, 0.4, 0.0}

\DeclareMathOperator*{\st}{subject~to}

\newtheorem{definition}{Definition}
\newtheorem{lemma}{Lemma}

\newtheorem{remark}{Remark}
\newtheorem{theorem}{Theorem}

\newcommand{\norm}[1]{\left\lVert#1\right\rVert}

\newcommand*{\genbf}[1]{\ifmmode\mathbf{#1}\else\textbf{#1}\fi}

\newcommand{\Phix}[0]{\bm{\Phi}^{\mathbf{x}}}
\newcommand{\Phiu}[0]{\bm{\Phi}^{\mathbf{u}}}
\newcommand{\Phig}[0]{\bm{\Phi}^{\! \bm{\nabla}}}
\newcommand{\Psix}[0]{\bm{\Psi}^{\mathbf{x}}}
\newcommand{\Psiu}[0]{\bm{\Psi}^{\mathbf{u}}}
\newcommand{\Psig}[0]{\bm{\Psi}^{\! \bm{\nabla}}}

\newcommand{\xb}[0]{\mathbf{x}}
\newcommand{\ub}[0]{\mathbf{u}}

\newcommand{\vb}[0]{\mathbf{v}}
\newcommand{\deltab}[0]{\bm{\delta}^{x_0}}
\newcommand{\pib}[0]{\bm{\pi}}

\usepackage{mathtools}

\DeclareMathOperator*{\argmax}{arg\,max}
\DeclareMathOperator*{\argmin}{arg\,min}

\definecolor{lightgray}{gray}{0.9}

\usepackage{tikz}
\usepackage{dsfont}

\begin{document}
\title{Learning to optimize with convergence guarantees using nonlinear system theory}

\author{Andrea Martin and Luca Furieri
\thanks{This research is supported by the Swiss National Science Foundation through the NCCR Automation (grant agreement 51NF40\textunderscore 80545) and the Ambizione grant PZ00P2\textunderscore208951.}
\thanks{A. Martin and L. Furieri are with the Institute of Mechanical Engineering, EPFL, Switzerland. E-mail addresses: \{andrea.martin,  luca.furieri\}@epfl.ch.}
}

\maketitle
\thispagestyle{empty} %

\begin{abstract}
The increasing reliance on numerical methods for controlling dynamical systems and training machine learning models underscores the need to devise algorithms that dependably and efficiently navigate complex optimization landscapes. Classical gradient descent methods offer strong theoretical guarantees for convex problems; however, they demand meticulous hyperparameter tuning for non-convex ones. The emerging paradigm of learning to optimize (L2O) automates the discovery of algorithms with optimized performance leveraging learning models and data -- yet, it lacks a theoretical framework to analyze convergence of the learned algorithms. In this paper, we fill this gap by harnessing nonlinear system theory. Specifically, we propose an unconstrained parametrization of all convergent algorithms for smooth non-convex objective functions. Notably, our framework is directly compatible with automatic differentiation tools, ensuring convergence by design while learning to optimize.
\end{abstract}


\section{Introduction}
\label{sec:introduction}
\IEEEPARstart{M}{any} fundamental tasks in machine learning (ML) and optimal control involve solving optimization problems, that is, computing a solution%
\begin{equation}
    \label{eq:objective_function}
    x^\star = \argmin\nolimits_{x \in \mathbb{R}^d} ~ f(x)\,,
\end{equation}
for a given objective function $f(\cdot)$. As the real-world applications of ML and optimal control grow in complexity, from training deep neural networks (NNs) for high-dimensional classification tasks to optimally operating large-scale cyber-physical systems, finding analytical solutions to these optimization problems becomes prohibitive. This has led to the increased use of numerical optimization algorithms, such as gradient descent methods, which iteratively approach the critical points $\hat{x}$ of $f(\cdot)$, i.e., the values $\hat{x}$ such that $\nabla \! f(\hat{x}) = 0$. As we rely on iterative algorithms to solve complex optimization problems, their ability to quickly and robustly converge to good critical points becomes crucial.

Traditionally, the optimization literature has focused on hand-crafting algorithms tailored to specific instances of \eqref{eq:objective_function}, such as those showcasing convex objective functions \cite{boyd2004convex}. For instance, widely-used optimization algorithms include vanilla gradient descent, the heavy-ball method, and Nesterov's accelerated method \cite{nesterov1983method}. While these algorithms come with strong theoretical guarantees for convex optimization, their performance on non-convex problems, such as training deep NNs, crucially depends on their hyperparameters \cite{bengio2012practical}. Besides, due to a lack of a general theory, hyperparameter tuning is often performed by domain experts based on best practices and know-how.

In the attempt to provide a unifying take on the analysis and synthesis of optimization algorithms, the control theory and ML communities have increasingly interpreted iterative update rules as evolving discrete-time dynamical systems; we refer to the recent review and road-map paper \cite{dorfler2024towards} for a comprehensive list of references. Notably, \cite{lessard2016analysis,lessard2022analysis,goujaud2023fundamental,scherer2021convex} have studied robustness and worst-case performance of optimization algorithms, leading to the design of new methods with optimized convergence rates \cite{lessard2016analysis,scherer2021convex} or sublinear regret guarantees in online optimization scenarios \cite{chen2023online}. All the results mentioned above are limited to convex objective functions. The paradigm of feedback optimization \cite{hauswirth2021optimization,belgioioso2022online} implements optimization algorithms directly in closed-loop with dynamical systems, endowing them with the ability to self-regulate and converge towards the solution of desired nonlinear optimization programs. Analyzing and shaping the transient performance of the resulting closed-loop behavior remains an open venue for research.

To tackle the non-convex and time-varying optimization landscapes that are ubiquitous in ML and optimal control, a learning to optimize (L2O) shift of paradigm has been emerging: moving from \emph{in silico} algorithms designed by hand based on general problem properties \cite{dorfler2024towards}, towards embracing ML to discover powerful algorithms from data. In this context, ``data'' refers to example optimization problems of interest provided during a training phase -- which can take place either offline or online. Specifically, the L2O approach parametrizes algorithms in a very general way, and performs meta-training over these parameters; we refer to \cite{chen2022learning} for a recent overview of L2O and a detailed discussion on its advantages and disadvantages. As observed in \cite{chen2022learning}, when the distribution of sample problems is narrow, learned algorithms can overfit the tasks and discover shortcuts that classic algorithms do not take. When the distribution of sample problems is sufficiently varied, the algorithm's performance transfers well to new tasks \cite{andrychowicz2016learning,li2017learning, li2017learning2}. However, it has been observed that optimizers trained as per \cite{andrychowicz2016learning} may lack convergence guarantees on almost all unseen tasks \cite{chen2022learning} -- even when they are taken from the same task distribution \cite{li2017learning2}. A mitigation to avoid compounding errors is proposed in \cite{li2017learning2} based on reinforcement meta-learning. For convex objective functions, provable convergence guarantees of learned optimizers were considered in \cite{heaton2023safeguarded} by exploiting a conservative fall-back mechanism that switches to a fixed convergent algorithm when the learned updates are too aggressive. To the best of the authors' knowledge, despite the outstanding empirical performance, the theoretical underpinnings of L2O such as convergence and robustness guarantees of the learned algorithms stand as uncharted territory.

\emph{Contributions:}
In this paper, we establish methods to learn high-performance optimization algorithms that are inherently convergent for smooth non-convex functions. From control system theory, we inherit the emphasis on convergence guarantees \cite{lessard2016analysis,scherer2021convex,hauswirth2021optimization,belgioioso2022online}, ensuring learned algorithms converge to local solutions in a provable and quantifiable way. From ML, we embrace the ability to tackle user-defined performance metrics through automatic differentiation\footnote{The computational engine to efficiently compute derivatives of functions specified by a computer program using the chain rule repeatedly, e.g., \cite{paszke2017automatic}.}, and the outstanding generalization capabilities to previously unseen optimization problems. Our key contribution is the reformulation of the problem of learning optimal convergent algorithms into an equivalent, unconstrained one that is directly amenable to automatic differentiation tools. 

We achieve this by dividing update rules into: $1)$ a gradient descent step that ensures convergence, and $2)$ a learnable term that enhances performance without compromising convergence. Notably, our method not only guarantees algorithm convergence, but it also encompasses \emph{all and only} convergent algorithms. As a result, we do not rely on safeguarding and early-stopping mechanisms \cite{andrychowicz2016learning,chen2022learning,heaton2023safeguarded}. Instead, the L2O approach we consider shifts the challenge to selecting a metric for algorithm performance that appropriately reflects the desiderata, for instance, trading off the speed of convergence and the quality of the solution. Furthermore, we achieve convergence even when dealing with incomplete gradient measurements, making our methodology relevant for ML with batch data. We validate the effectiveness and generalizability of our methodology through ML benchmarks.%

\emph{Notation:} The set of all sequences $\mathbf{x} = (x_0,x_1,x_2,\ldots)$ where $x_t \in \mathbb{R}^n$ for all $t\in \mathbb{N}$ is denoted as $\ell^n$.  For $\mathbf{x} \in \ell^n$, we denote by $z\mathbf{x} = (x_1,x_2,\ldots)$ the sequence shifted one-time-step forward.  Moreover,  $\mathbf{x}$ belongs to $\ell_2^n \subset \ell^n$ if $\norm{\mathbf{x}}_2 = \sqrt{\sum_{t=0}^\infty |x_t|^2} < \infty$, where $|\cdot|$ denotes any vector norm. When clear from the context, we omit the superscript $n$ from $\ell^n$ and $\ell^n_2$. For a function $g:\mathbb{R}^n \rightarrow \mathbb{R}^m$, we write $g\left(\mathbf{x}\right) = (g(x_0),g(x_1),\ldots) \in \ell^m$. A causal operator $\mathbf{A}:\ell^n \rightarrow \ell^m$ such that $\mathbf{A}(\mathbf{x}) = (A_0(x_0),A_1(x_{1:0}),\ldots,A_t(x_{t:0}),\ldots)$ is said to be $\ell_2$-stable if $\mathbf{A}(\mathbf{x}) \in \ell_{2}^m$ for all $\mathbf{x} \in \ell_{2}^n$.  Equivalently, we  write $\mathbf{A} \in \mathcal{L}_{2}$. We denote by $\left \lfloor x \right \rfloor$ the greatest integer smaller than $x\in \mathbb{R}$ and use $a \bmod b$ to denote the remainder of $a \in \mathbb{N}$ when divided by $b \in \mathbb{N}$.

\section{Problem Formulation}
In this paper, we focus on optimization problems in the form \eqref{eq:objective_function} where $f(\cdot)$ has $\beta$-Lipschitz gradients, that is, $|\nabla \! f(x) - \nabla \! f(y)| \leq \beta |x - y|$ for all $x, y \in \mathbb{R}^d$.
We denote the set of such $\beta$-smooth functions by $\mathcal{S}_\beta$. Further, it is assumed that $f(\cdot)$ is bounded from below. %
 We describe an iterative optimization algorithm via the recursion
\begin{equation}
    \label{eq:algorithm_dynamics}
    x_{t+1} = x_t + u_t = x_t + \pi_t(f,x_{t:0})\,, \quad t \in \mathbb{N}\,,
\end{equation}
where $x_0 \in \mathbb{R}^d$ is the initial guess, $x_t \in \mathbb{R}^d$ is the candidate solution vector after $t$ iterations, and $u_t = \pi_t(f,x_{t:0}) \in \mathbb{R}^d$ is the algorithm update rule. We can write \eqref{eq:algorithm_dynamics} compactly as
\begin{equation}
\label{eq:algorithm_dynamics_operator_form}
    z \xb = \xb + \pib(f,\xb) + z \deltab\,,
\end{equation}
where $\bm{\pi}(f,\cdot) = \left(\pi_0(f,x_0),\pi_1(f,x_{1:0}),\ldots\right)$ is a causal operator for any objective function $f$. The initial state sequence $\bm{\delta}^{x_0}$ is defined as $\deltab = (x_0, 0, \ldots) \in \ell_2$.  We proceed to define the fundamental notion of convergent algorithms.
\begin{definition}
    \label{def:convergent_alg}
    Consider the iteration \eqref{eq:algorithm_dynamics}. An update rule $\bm{\pi}(f,\mathbf{x})$ is \emph{convergent} for $f$ if for any $x_0 \in \mathbb{R}^d$
    \begin{equation}
    \label{eq:asymptotic_convergence}
       \lim_{t\rightarrow \infty} \pi_t(f, x_{t:0}) = 0, \quad \lim_{t \rightarrow \infty} \nabla \! f(x_t) = 0\,.
    \end{equation}
Equivalently, we write \mbox{$\bm{\pi} \in \Gamma(f)$}. Additionally, if
    \begin{align}
    \label{eq:square_sum_convergence}
         \norm{\bm{\pi}(f,\mathbf{x})}^2  < \infty \,, \quad 
         \norm{\nabla \! f(\mathbf{x})}^2  < \infty \,, %
    \end{align}
    we say the algorithm is \emph{square-sum convergent} for $f$. Equivalently, we write \mbox{$\bm{\pi} \in \Sigma(f)$}.
\end{definition}

Note that every update rule in $\Sigma(f)$ also lies in $\Gamma(f)$ for every $f \in \mathcal{S}_\beta$. In particular, although \eqref{eq:asymptotic_convergence} and \eqref{eq:square_sum_convergence} both guarantee convergence to a critical point of $f$ as $t \rightarrow \infty$,  \eqref{eq:square_sum_convergence} only holds for those algorithms in $\Gamma(f)$ that achieve a sufficiently fast asymptotic convergence rate. Further, we observe that classical convergence bounds for smooth convex optimization in the form $|x_{t}-x^\star|\leq K \rho^t|x_0-x^\star|$, where $K>0$, see, e.g, \cite{lessard2016analysis,scherer2021convex}, readily imply that $\nabla \! f(\mathbf{x}) \in \ell_2$.

Given a distribution $\mathcal{F}$ over functions in $\mathcal{S}_\beta$ and a distribution $\mathcal{X}_0$ over initial guesses $x_0 \in \mathbb{R}^d$, the problem of designing an optimal convergent algorithm is formulated as
\begin{subequations}
\label{eq:metaopt_constrained_problem}
    \begin{alignat}{3}
   &~\min_{\bm{\pi}} && \mathbb{E}_{f \sim \mathcal{F}, x_0 \sim \mathcal{X}_0} \left[ \mathtt{MetaLoss}(f, \xb) \right] \label{eq:expected_meta_loss} \\%\mathtt{MetaLoss}(\mathcal{F},\mathcal{X}_0, \xb) \\
   &\st && \quad x_{t+1} = x_t + \pi_t(f,x_{t:0}),\label{eq:algo_def_metaloss_problem} \\
   &~&& \quad \bm{\pi}(f,\mathbf{x}) \in \Sigma(f),~ \forall f \in \mathcal{S}_\beta\label{eq:constraint_convergence}\,,
\end{alignat}
\end{subequations}
where $\Sigma(f)$ can be relaxed to $\Gamma(f)$ depending on the design specifications. As suggested in \cite{andrychowicz2016learning,li2017learning, li2017learning2}, a useful choice for $\mathtt{MetaLoss}(f, \xb)$ in \eqref{eq:expected_meta_loss} is given by %
\begin{equation}
\label{eq:metaobjective_mathematical_expression}
    \mathtt{MetaLoss} (f,\xb) = \sum\nolimits_{t=0}^T \alpha_t |\nabla \! f(x_t)|^2+ \gamma_t f(x_t)\,,
\end{equation}
where $\alpha_t \geq 0$ and $\gamma_t \geq 0$. Specifically, $\alpha_t$ promotes fast convergence of $\nabla \! f(x_t)$, while $\gamma_t$ drives the algorithm closer to the solution of \eqref{eq:objective_function}. As a result, $\alpha_t$ and $\gamma_t$ act as hyperparameters that must be tuned to strike a balance between these two competing aspects.
At the same time, the constraint \eqref{eq:constraint_convergence} ensures convergence to a critical point $\hat{x}$ satisfying $\nabla \! f(\hat{x})=0$ for any future $f \in \mathcal{S}_\beta$. 
\begin{remark}[The value of convergence] Excluding update rules that fail to comply with \eqref{eq:constraint_convergence} is crucial, as \eqref{eq:constraint_convergence} ensures algorithm convergence, even if meta-optimization is prematurely stopped. Notably, convergence is necessary for generalizing to unseen problems $f_{\operatorname{new}} \in \mathcal{S}_{\beta}$ drawn from a different distribution $\mathcal{F}^{\prime}$ and achieving sublinear meta-regret in online convex optimization \cite{chen2023online}. %

\end{remark}

\smallskip

\section{Main Results}

This section characterizes update rules that converge according to Definition~\ref{def:convergent_alg}, and describes how to learn over them. First, given full gradient measurements, we establish a complete parametrization of all and only the algorithms that converge in the sum-square sense as per \eqref{eq:constraint_convergence}. Second, for the case -- common in ML and deep learning applications -- where $f(x) = \sum_{i=0}^{M-1} f_i(x)$ and only partial gradients $\nabla \! f_i(x)$ are available at each step, we parametrize algorithms that converge asymptotically as per \eqref{eq:asymptotic_convergence}. In both scenarios, we directly parametrize convergent update rules via a vector $\theta \in \mathbb{R}^D$, thus enabling unconstrained learning of convergent-by-design algorithms via automatic differentiation tools.

\subsection{Learning over all square-sum convergent algorithms}
We start by proving that any update rule in the form
\begin{equation}
\label{eq:separation}
    \bm{\pi}(f, \mathbf{x}) = - \eta \nabla \! f(\mathbf{x}) + \mathbf{v}\,, 
\end{equation}
lies in $\Sigma(f)$ for any $\mathbf{v} \in \ell_2$ and any $f \in \mathcal{S}_\beta$, as long as $0 < \eta < \beta^{-1}$. In other words, if we perturb standard gradient descent with an $\ell_2$ \emph{``enhancement''} term -- designed, e.g., to escape a bad local minimum or a saddle point --  we preserve square-sum convergence to a critical point of $f$.

\begin{lemma}
\label{le:sufficiency}
    Consider the recursion $\eqref{eq:algorithm_dynamics_operator_form}$. The update rule given by \eqref{eq:separation} with $0 < \eta <\beta^{-1}$ satisfies \eqref{eq:constraint_convergence} for every choice of $\mathbf{v} \in \ell_2$.
\end{lemma}

The class of algorithms in the form \eqref{eq:separation} suggests a useful separation of roles; a gradient descent update can be used to ensure convergence, while an enhancement term $\mathbf{v} \in \ell_2$ can be learned to improve the algorithm performance. %
Nonetheless, a crucial question regarding the conservatism of searching over $\mathbf{v} \in \ell_2$ in %
\eqref{eq:separation} remains.

\emph{Can \emph{any} convergent algorithm complying with \eqref{eq:constraint_convergence} be written as the sum of a gradient-based update and an enhancement signal $\mathbf{v} \in \ell_2$ as per \eqref{eq:separation}?}

In what follows, we answer in the affirmative,  further revealing that $\mathbf{v} \in \ell_2$ must be parametrized as a function of $\deltab$ in order to recover any convergent behavior using \eqref{eq:separation}. Our proof hinges on studying the closed-loop mappings induced by an update rule $\bm{\pi}(f, \xb)$.%
\begin{definition}
Consider the recursion $\eqref{eq:algorithm_dynamics_operator_form}$. For any update rule $\bm{\pi}(f,\mathbf{x})$, the mapping $(f,\deltab){\rightarrow} (\mathbf{x},\mathbf{u},\nabla \! f(\mathbf{x}))$ is denoted as the \emph{closed-loop mapping} induced by $\mathbf{u} = \bm{\pi}(f,\mathbf{x})$.%
\end{definition}

The terminology above is drawn from control system theory. Under a system-theoretic lens, we can view $\bm{\pi}(f, \xb)$ as an objective-dependent state feedback control policy, and $(f,\deltab){\rightarrow} (\mathbf{x},\mathbf{u},\nabla \! f(\mathbf{x}))$ as the corresponding closed-loop behavior. The constraint $\eqref{eq:constraint_convergence}$ thus translates to regulating the system output signal $\mathbf{y}$ to $0$, further requiring  that $\mathbf{y}=\nabla \! f(\xb)\in \ell_2$, robustly for any $x_0 \in \mathbb{R}^d$ and any $f \in \mathcal{S}_\beta$. 
\begin{lemma}
    \label{le:necessity}
    Let $x_0 \in \mathbb{R}^d$ and $f \in \mathcal{S}_\beta$. Define
    \begin{equation}
    \label{eq:induced_by_pi}
        (f,\bm{\delta}^{x_0}) \rightarrow (\xb_{\pib}, \ub_{\pib} ,\nabla \! f(\mathbf{x}_{\pib}))\,,  
    \end{equation} 
    as the closed-loop mapping induced by a policy $\mathbf{u} = \bm{\pi}(f,\mathbf{x})$. For any $\bm{\pi}(f,\mathbf{x})$ complying with \eqref{eq:constraint_convergence}, there exists an operator $\mathbf{V} \in \mathcal{L}_2$ such that the closed-loop mapping given by
    \begin{equation}
    \label{eq:V_CL}
       (f,\bm{\delta}^{x_0}) \rightarrow (\mathbf{x}, -\eta \nabla \! f(\xb) + \mathbf{V}(\bm{\delta}^{x_0}), \nabla \! f(\mathbf{x}))\,,
    \end{equation} 
     is equivalent to \eqref{eq:induced_by_pi}.
\end{lemma}

The completeness property stated above is key, as it implies that \eqref{eq:separation} encompasses all sum-square convergent algorithms -- including those that \emph{globally} minimize \eqref{eq:expected_meta_loss}. Together with Lemma~\ref{le:sufficiency}, %
Lemma~\ref{le:necessity} leads to our main result.  %
\begin{theorem}
\label{th:reformulation}
    If $0 < \eta < \beta^{-1}$, the meta-optimization problem \eqref{eq:metaopt_constrained_problem} is equivalent to
    \begin{subequations}
    \label{eq:metaopt_constrained_problem_L2_operator}
    \begin{alignat}{3}
   & \min_{\mathbf{V} \in \mathcal{L}_2} && \mathbb{E}_{f \sim \mathcal{F}, x_0 \sim \mathcal{X}_0} \left[ \mathtt{MetaLoss}(f, \xb) \right] 
    \\
   &\st && \quad z\mathbf{x} = \mathbf{x}-\eta \nabla \! f(\mathbf{x})+\mathbf{V}(\bm{\delta}^{x_0})+z\bm{\delta}^{x_0} \label{eq:metaopt_constrained_problem_L2_operator_dynamics}\,.
\end{alignat}
\end{subequations}
\end{theorem}

\smallskip

A few comments are in order. First, any possibly suboptimal solution $\mathbf{V} \in \mathcal{L}_2$ to \eqref{eq:metaopt_constrained_problem_L2_operator} yields a converging algorithm complying with \eqref{eq:constraint_convergence}. Second, every converging algorithm complying with \eqref{eq:constraint_convergence} is recovered by appropriately choosing $\mathbf{V} \in \mathcal{L}_2$ with no conservatism. Third, as the convergence constraint \eqref{eq:constraint_convergence} simplifies to $\mathbf{V} \in \mathcal{L}_2$, we can use finite-dimensional approximations of operators in $\mathcal{L}_2$, that is
\begin{equation}
    \label{eq:V:stable}
    \mathbf{V}(\bm{\delta}^{x_0}, \theta) \in \ell_2, \quad \forall \bm{\delta}^{x_0}\in \ell_2, ~\forall \theta \in \mathbb{R}^D\,,
\end{equation}
to translate \eqref{eq:metaopt_constrained_problem_L2_operator} into the unconstrained\footnote{Note that \eqref{eq:metaopt_constrained_problem_L2_operator_dynamics} defines the signal $\mathbf{x}$ through the recursion \eqref{eq:algorithm_dynamics} and does not pose any constraints on $\mathbf{V} \in \mathcal{L}_2$.} optimization problem of learning the best parameter $\theta\in \mathbb{R}^D$. To ensure that \eqref{eq:V:stable} holds, one can, for instance, model $\mathbf{V}$ as a stable recurrent NN 
$v_t = \phi(\theta,v_{t-1},u_t)$, where $\phi(\cdot)$ is contracting for all $\theta \in \mathbb{R}^D$; several such models have recently been developed in the literature \cite{kim2018standard,revay2023recurrent} and are readily implementable. Despite approximating the original infinite-dimensional problem \eqref{eq:metaopt_constrained_problem}, these parametrizations have been shown to be highly expressive \cite{revay2023recurrent}, with formal density and suboptimality bounds for linear operators in $\mathcal{L}_2$ \cite{fisher2023approximation}.

In practice, it may prove beneficial to introduce explicit dependence of $\mathbf{V}$ in \eqref{eq:V:stable} on additional input features besides $\deltab$. Indeed, as shown in \cite{li2017learning2}, learning over algorithms that react to $(x_{t:0},\nabla \! f(x_{t:0}),f(x_{t:0}))$ can be effective in transferring their meta-performance to ML tasks vastly different from those encountered during training. While Theorem~\ref{th:reformulation} proves that designing an update rule that solely reacts to $x_0\sim\mathcal{X}_0$ is sufficient for achieving meta-optimal behaviors, additional input features could significantly improve how effectively we navigate the meta-optimization landscape. For instance, by defining $\bm{\omega} = \bm{\Omega}(\mathbf{x},\nabla \! f(\mathbf{x}),f(\mathbf{x}))$ and $\mathbf{z} = \mathbf{Z}(\deltab)$, where $\bm{\Omega}:\ell\rightarrow \ell$ and $\mathbf{Z} \in \mathcal{L}_2$ are operators to be freely designed, we can generate $\mathbf{v} \in \ell_2$ as follows %
\begin{equation}
\label{eq:input_features}
    v_t = |z_t||\omega_t|^{-1} \omega_t\,.%
\end{equation}
Using \eqref{eq:input_features}, sum-square convergence is preserved by design, as we set $|v_t| = |z_t|$ at all times, and $\mathbf{z} \in \ell_2$. Further, completeness as per Lemma~\ref{le:necessity} is maintained; this is proved by choosing $\mathbf{Z}$ according to \eqref{eq:choice_of_V} in the Appendix and selecting $\bm{\Omega}(\mathbf{x},\nabla \! f(\mathbf{x}),f(\mathbf{x})) = \mathbf{Z}(\bm{\delta}^{x_0})$. We remark that, even though completeness is guaranteed, discovering alternative parametrizations of converging algorithms beyond \eqref{eq:separation} with \eqref{eq:input_features} could be beneficial; indeed, despite their theoretical equivalence, different convergence strategies may result in more favorable meta-optimization landscapes.

\subsection{The case of gradients with errors}
In many ML and deep learning tasks, $f(x)$ is obtained as the empirical average of the cost over a batch of input data, that is, $f(x) = \sum_{i=0}^{M-1} f_i(x)$. In these cases, global gradient information $\nabla \! f(x)$ may not be available, and the candidate solution $x_t$ is updated based on $\nabla \! f_i(x_t)$ for some $i \in [0, M-1]$ only. Drawing connections with analysis techniques for stochastic gradient descent (SGD) from \cite{bertsekas2000gradient}, we proceed to parametrize a rich class of asymptotically convergent algorithms that rely on partial gradient information.

\begin{theorem}
   \label{th:sufficiency_batch}
    Let $f(x) = \sum_{i=0}^{M-1} f_i(x)$ be separable in $M \in \mathbb{N}$ continuously differentiable components $f_i \in \mathcal{S}_\beta$ satisfying, for some non-negative constants $A$ and $B$,%
    \begin{equation}
    \label{eq:assumption_batch_gradients}
        |\nabla \! f_i (x)| \leq A + B |\nabla \! f (x)|\,, ~\forall x \in \mathbb{R}^d\,.\footnotemark %
    \end{equation}
    \footnotetext{This assumption encompasses the case of Lipschitz continuous function components $f_i$ when $B=0$, and is therefore common in the analysis of SGD-related methods \cite{bertsekas2000gradient}.}Choose any stepsize sequence $\bm{\eta} \in \ell_2$ such that $\sum_{t = 0}^{\infty} \eta_t = \infty$ with $\eta_t > 0$ at all times, and any $\mathbf{v}$ such that
    \begin{equation}
        \label{eq:assumption_v}
        |v_t| \leq \eta_{\left \lfloor t/M \right \rfloor} (C + D |\nabla \! f (x_t)|)\,,
    \end{equation}
    for some non-negative constants $C$ and $D$. Then, the update  %
    \begin{align}
        \label{eq:update_batch}
       \pi_t(f,x_t) = -\eta_{\left \lfloor t/M \right \rfloor} (\nabla \! f_{t \bmod M}(x_t) + v_t)\,,\footnotemark
    \end{align}
    \footnotetext{Unlike SGD that selects partial gradients $\nabla f_{k_t}$ with $k_t$ drawn uniformly at random from $\{0,1,\ldots,M-1\}$,  our analysis considers the case where $k_t$ is deterministically selected in a sequential way, thus resulting in a deterministic convergence result.}is convergent according to \eqref{eq:asymptotic_convergence}, that is, $\bm{\pi}(f,\mathbf{x}) \in \Gamma(f)$.
    
\end{theorem}

The proof of Theorem~\ref{th:sufficiency_batch} adapts the analysis of \cite[Proposition~2]{bertsekas2000gradient} by accounting for the contribution of $v_t$ as per \eqref{eq:update_batch}. To ensure \eqref{eq:assumption_v}, while endowing our learned algorithms with the ability to react to input features, we propose using
\begin{align}
& v_t = \eta_{\left \lfloor t/M \right \rfloor} |z_t| |\omega_t|^{-1} \omega_t\,, \quad\mathbf{z} = \mathbf{Z}(\deltab) \label{eq:implementation_batch}\,,\\
&\omega_t = \Omega_t(x_{t:0},\nabla\!f_{\tau}(x_{t}), \ldots,\nabla\!f_{0}(x_{0}),f_{\tau}(x_{t}),\ldots, f_0(x_0))\,, \nonumber
\end{align}
where $\tau = t \bmod M$, and $\bm{\Omega}:\ell\rightarrow \ell$ and $\mathbf{Z} \in \mathcal{L}_2$ are operators to be freely designed. 
In this way, similarly to \eqref{eq:input_features}, we have $|v_t| = \eta_{\left \lfloor t/M \right \rfloor}|z_t|\leq\eta_{\left \lfloor t/M \right \rfloor} \max_{t \in \mathbb{N}} ~ z_t$, and we thus satisfy \eqref{eq:assumption_v} with $C= \max_{t \in \mathbb{N}} ~ z_t$ and $D = 0$.

The update rule \eqref{eq:update_batch} cycles through the gradients $\nabla \! f_i$ and applies an enhancement signal to be learned. Coherently with standard SGD, Theorem~\ref{th:sufficiency_batch} guarantees asymptotic convergence of the learned algorithm -- despite the additional presence of $\vb$ satisfying \eqref{eq:assumption_v}. Moreover, while \eqref{eq:assumption_v} may restrict the set of $\pib(f, \xb) \in \Gamma(f)$ that our parametrization can achieve, we proceed to illustrate the rich expressivity of learning over enhancement terms modeled as \eqref{eq:implementation_batch}.

\section{Experiments}
Motivated by \cite{andrychowicz2016learning}, we consider the problem of learning to optimize the parameters $x_t$ of a shallow NN for image classification with the MNIST dataset. Further, we investigate how our optimizer generalizes to different network activation functions and different initial parameter distribution $\mathcal{X}_0$.

We model our trainable optimizer as per \eqref{eq:update_batch}-\eqref{eq:implementation_batch}, using a recurrent equilibrium network\footnote{This architecture, which subsumes several existing deep NN models, proves convenient as it provides a finite-dimensional approximation of $\mathcal{L}_2$ without imposing any constraint on the parameter vector $\theta$.} \cite{revay2023recurrent}  with depth of $r = 3$ layers and internal state dimension $n = 3$ as a model for $\mathbf{Z}(\theta) \in \mathcal{L}_2$, and a multilayer perceptron (MLP) with $2$ hidden layers as a model for $\bm{\Omega}(\theta)$. We instead model the shallow NN as a simple perceptron that, given a vectorized image $s$ corresponding to a handwritten digit as input, predicts the label $\hat{l}(s, x_t)$ according to the criterion:%
\begin{equation}
    \label{eq:classifier_output_prediction}
    \argmax\nolimits_{i} ~ o(s, x_t) = \argmax\nolimits_{i} ~ [\operatorname{tanh}(s W_t^\top + b_t)]_{i}\,,
\end{equation}
where $W_t$ and $b_t$ are the trainable parameters of the classifier, whose scalar entries are collected in $x_t$, and $[\cdot]_{i}$ denotes the $i$-th entry of a vector. Based on \eqref{eq:classifier_output_prediction}, we also define the classification loss $g(s, l, x_t)$ on an image $s$ as the cross entropy loss between the softmax transformation of $o(s, x_t)$ and $l$, the true label of $s$, encoded as a one-hot vector.

To encourage our learned updates \eqref{eq:update_batch} to promote the accuracy of the classifier predictions \eqref{eq:classifier_output_prediction} after $T = 50$ iterations of \eqref{eq:algorithm_dynamics}, we consider the meta-loss \eqref{eq:metaobjective_mathematical_expression} with $\alpha_t = 0$, $\gamma_t = 0.95^{T-t}$, and let $f(x_t)$ be the cross entropy loss on the training dataset. For fixed parameters $\theta$ of $\mathbf{Z}$ and $\bm{\Omega}$, we approximate \eqref{eq:expected_meta_loss} by repeating the training of $W_t$ and $b_t$ in \eqref{eq:classifier_output_prediction} for $10$ times, using different initial parameters sampled from a distribution $\mathcal{X}_0$ that is uniform in the interval $[0, 0.01]$. Motivated by Theorem~\ref{th:sufficiency_batch}, we estimate $f(x_t)$ and $\nabla \! f (x_t)$ in \eqref{eq:metaopt_constrained_problem_L2_operator_dynamics} using random minibatches of $128$ images drawn sequentially. Then, consistently with \cite{andrychowicz2016learning}, we perform the minimization of \eqref{eq:expected_meta_loss} using Adam with a learning rate of $0.01$. We use $80\%$ of the training dataset for optimizing $\theta$. 

After $40$ training epochs, we freeze the value of $\theta$, and we benchmark the performance of our learned optimizer against standard optimizers including Adam, SGD, Nesterov’s accelerated gradient (NAG), and RMSprop. To assess the generalization capabilities of \eqref{eq:implementation_batch}, we use the remaining $20\%$ of the training dataset to train a shallow NN that: $1)$ uses $\operatorname{sigmoid}$ or $\operatorname{ReLU}$  as activations in \eqref{eq:classifier_output_prediction}, and $2)$ whose initial parameters $[x_0]_i$ are sampled from independent and identically distributed Gaussian distributions $\mathcal{N}(0, 0.1)$.%

\begin{figure*}[htb]
    \centering
    \subfloat[Activation function: $\operatorname{tanh}$. \label{fig:init_uniform}]{\includegraphics[width = 0.682\columnwidth]{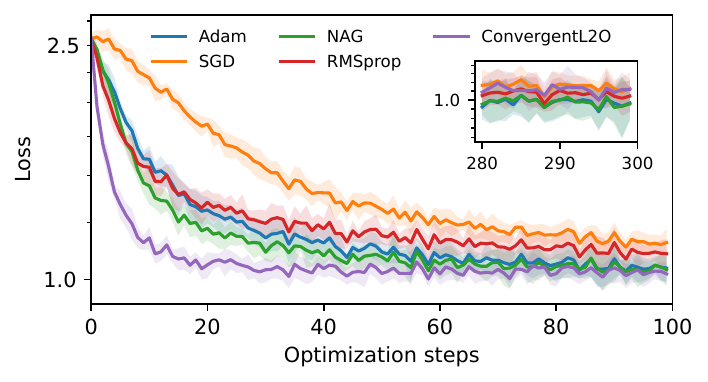}}
    \hfill
    \subfloat[Activation function: $\operatorname{sigmoid}$. \label{fig:init_normal}]{\includegraphics[width = 0.682\columnwidth]{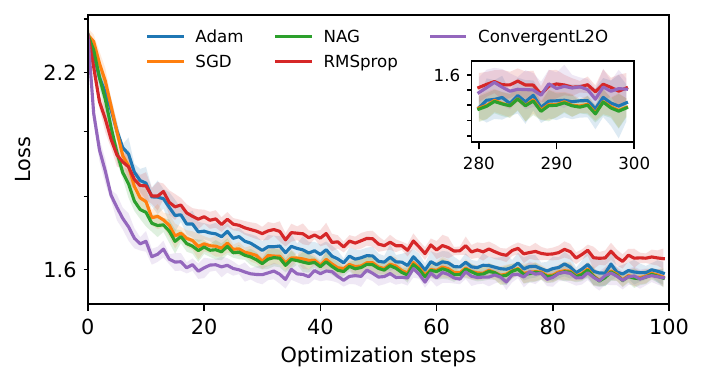}}
    \hfill
    \subfloat[Activation function: $\operatorname{ReLU}$. \label{fig:init_normal}]{\includegraphics[width = 0.682\columnwidth]{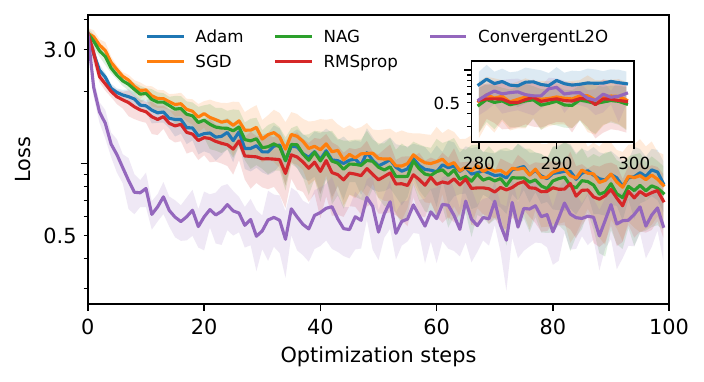}}
  \caption{Training curves of learned and hand-crafted optimizers; shaded areas and solid lines denote standard deviations and mean values, respectively.}
  \label{fig:comparison_learned_hand_crafted_optimzers}
\end{figure*}

We report training curves for our learned optimizer and for classical hand-crafted algorithms in Figure~\ref{fig:comparison_learned_hand_crafted_optimzers} and the corresponding average test accuracy in the tables below.\footnote{Following \cite{andrychowicz2016learning}, for each considered scenario, we tune the learning rate of each baseline optimizer to minimize the training loss $f(x_t)$, and we adopt default values for additional hyperparameters.} 
\begin{table}[ht]
    \centering
    \rowcolors{1}{}{lightgray}
    \begin{tabular}{c|ccc}
        Step $t = 20$ & $\operatorname{tanh}$ & $\operatorname{sigmoid}$ & $\operatorname{ReLU}$\\
        \hline
        Adam & $71.7 \pm 5.1\%$ & $76.1 \pm 3.1\%$ & $52.7 \pm 11.1\%$\\
        SGD & $44.9 \pm 4.2\%$ & $79.7 \pm 1.9\%$& $49.8 \pm 9.3\%$\\
        NAG & $79.7 \pm 1.4\%$ & $81.1 \pm 1.5\%$ & $52.7 \pm 10.2\%$\\
        RMSprop & $69.4 \pm 2.9\%$ & $72.8 \pm 2.3\%$ & $61.1 \pm 8.9\%$\\
        ConvergentL2O & \color{darkgreen}$\mathbf{87.0 \pm 0.5}\%$ &\color{darkgreen} $\mathbf{86.8 \pm 0.6}\%$ & $86.3 \pm 0.6\%$\\
        LSTM & $82.2 \pm 0.1\%$ & $83.3 \pm 0.1\%$ & \color{darkgreen} $\mathbf{88.3 \pm 0.0}\%$\\
        \hline
     \hline
    \end{tabular}
    \vspace{0.25cm}\\
    \centering
    \rowcolors{1}{}{lightgray}
    \begin{tabular}{c|ccc}
        Step $t = 300$ & $\operatorname{tanh}$ & $\operatorname{sigmoid}$ & $\operatorname{ReLU}$\\
        \hline
        Adam & \color{darkgreen} $\mathbf{89.5 \pm 0.5}\%$ & \color{darkgreen} $\mathbf{89.6 \pm 0.3}\%$ & $70.3 \pm 12.2\%$\\
        SGD & $87.4 \pm 0.4\%$ & $89.3 \pm 0.3\%$ & $80.6 \pm 8.1\%$\\
        NAG & $89.4 \pm 0.2\%$ & $89.4 \pm 0.2\%$ & $82.2 \pm 7.6\%$\\
        RMSprop & $87.6 \pm 2.1\%$ & $88.5 \pm 0.4\%$ & $81.5 \pm 7.5\%$\\
        ConvergentL2O & $88.5 \pm 0.2\%$ & $88.4 \pm 0.3\%$ & $87.7 \pm 0.2\%$\\
        LSTM & $81.4 \pm 0.0\%$ & $81.4 \pm 0.0\%$ & \color{darkgreen} $\mathbf{88.3 \pm 0.0}\%$\\
        \hline
        \hline
    \end{tabular}
\end{table}
In all considered scenarios, learned algorithms excel in finding shortcuts, that is, in steering $x_0$ to a good local minimum within a few iterations; this is reflected in the superior test accuracy achieved after only $t=20$ optimization steps.  As the gradient norm diminishes, our learned algorithm favors simple gradient-based updates to ensure convergence, as predicted by Theorem~\ref{th:sufficiency_batch}. Despite being trained to optimize for $T = 50$ optimization steps only, the average test accuracy of our method matches that of classical algorithms after $300$ iterations of \eqref{eq:algorithm_dynamics} -- compare also the zooming at $t = 300$ in Figure~\ref{fig:comparison_learned_hand_crafted_optimzers}. Remarkably, our algorithm generalizes well also to the optimization landscape of the $\operatorname{ReLU}$ classifier, which may prove particularly challenging, as highlighted in \cite{andrychowicz2016learning}, due to its structural difference with respect to $\operatorname{tanh}$ in \eqref{eq:classifier_output_prediction}. Future work will address the generalization of algorithms trained on the MNIST dataset to different test datasets, e.g., Fashion-MNIST; such generalization was not achieved using the current shallow classifier architecture \eqref{eq:classifier_output_prediction}.

To compare with alternative L2O approaches  \cite{andrychowicz2016learning}, we have trained for $200$ epochs a two-layer Long Short-Term Memory (LSTM) optimizer $u_t = \operatorname{LSTM}(x_{t}, \nabla \! f(x_t), f(x_t))$. As shown in the table above, the LSTM optimizer achieves similar average test accuracy as our $\operatorname{ConvergentL2O}$ algorithm. Nonetheless, the LSTM output $u_t$ does not vanish with time, causing the classifier parameters to diverge\footnote{We refer to \href{https://github.com/andrea-martin/ConvergentL2O}{https://github.com/andrea-martin/ConvergentL2O} for the corresponding plots and the source code reproducing our numerical examples.} -- similar phenomena were also observed in \cite{li2017learning2}. None of our simulations exhibited such divergence as per Theorem~\ref{th:sufficiency_batch}.

\section{Conclusion}
In this paper, we have introduced a methodology for learning over all convergent update rules for smooth non-convex optimization, thus enabling the automated synthesis of more reliable, efficient, and reconfigurable algorithms. By synergizing nonlinear system theory with the emerging L2O paradigm, we aimed to close the gap between offline, theory-based algorithm design and adaptable, example-driven approaches that are the hallmark of ML. Building on the proposed control-theoretic perspective we have embraced, further avenues for future research include certifying stronger convergence guarantees of learned algorithms for convex optimization,  analyzing generalization capabilities,  extending our framework to online and constrained optimization scenarios, and federated learning.

\bibliographystyle{IEEEtran}
\bibliography{references}

\appendix

\begin{proof}[Lemma~\ref{le:sufficiency}]
    Let $\nabla \! f (x_t) = \nabla_{\! t}$ for compactness. For any $t \in \mathbb{N}$ and $f \in \mathcal{S}_\beta$, it holds that $f(x_{t+1}) \leq f(x_t) + \nabla_{\! t}^\top (x_{t+1} - x_t) + \frac{\beta}{2} |x_{t+1} - x_t|^2$.
    Substituting $x_{t+1} - x_t = v_t - \eta  \nabla_{\! t}$ according to \eqref{eq:separation} yields
    \begin{equation}
        \label{eq:proof_gd_sufficiency_intermediate}
        f(x_{t+1}) \leq f(x_t) - \eta |\nabla_{\! t}|^2 + \nabla_{\! t}^\top v_t  + \frac{\beta}{2} |v_t - \eta \nabla_{\! t}|^2\,.
    \end{equation}
    Observe that for any $a, b \in \mathbb{R}^d$ and any $\epsilon > 0$, we have that $a^\top b \leq |a| |b| \leq \frac{|a|^2}{2 \epsilon} + \frac{\epsilon |b|}{2}$ and $\frac{1}{2} |a - b|^2 \leq |a|^2 + |b|^2$ thanks to the Cauchy-Schwarz and Young's inequalities.
    Hence, we can upper-bound the right-hand side of \eqref{eq:proof_gd_sufficiency_intermediate} by $f(x_t) - \eta |\nabla_{\! t}|^2 + \frac{|\nabla_{\! t}|^2}{2 \epsilon} + \frac{\epsilon |v_t|^2}{2} + \beta \left(|v_t|^2 + \eta^2 |\nabla_{\! t}|^2 \right)$.
    Collecting terms and letting $\rho = 2 \eta \epsilon (1 - \beta \eta) - 1$, we obtain
    \begin{equation}
        \label{eq:proof_gd_sufficiency_intermediate_2}
        \frac{\rho}{2 \epsilon} |\nabla_{\! t}|^2 \leq f(x_t) - f(x_{t+1}) + \left(\frac{\epsilon}{2} + \beta \right) |v_t|^2\,.
    \end{equation}
    As $1 - \beta \eta > 0$, choosing any $\epsilon > \frac{1}{2 \eta (1 - \beta \eta)} > 0$ ensures that $\rho > 0$. Then, summing \eqref{eq:proof_gd_sufficiency_intermediate_2} with $t$ ranging from $0$ to $T \in \mathbb{N}$ and observing that the term $f(x_t) - f(x_{t+1})$ telescopes and that $f(x_T) \geq \inf_{x \in \mathbb{R}^d} f(x)$, we obtain $\sum_{t = 0}^{T} |\nabla_{\! t}|^2 \leq \frac{2 \epsilon}{\rho}(f(x_0) - \inf_{x \in \mathbb{R}^d} f(x)) + \frac{\epsilon}{\rho} \left(\epsilon + 2\beta \right) \sum_{t = 0}^{T} |v_t|^2$.
    As $f$ is bounded from below, $f(x_0) - \inf_{x \in \mathbb{R}^d} f(x)$ is finite and non-negative. Finally, as $\vb \in \ell_2$, taking the limit of $T$ to $\infty$ yields $\norm{\nabla \! f(\xb)}^2 < \infty$, which concludes the proof.
\end{proof}

\begin{proof}[Lemma~\ref{le:necessity}]
    For any $f \in \mathcal{S}_\beta$ and any $\bm{\pi}(f, \xb)$ complying with \eqref{eq:constraint_convergence}, select the operator $\mathbf{V}$ as
    \begin{equation}
    \label{eq:choice_of_V}
        \mathbf{V}(\deltab) = \eta \nabla \! f(\mathbf{x}_{\bm{\pi}})+\mathbf{u}_{\bm{\pi}}\,.
    \end{equation}
    As $\bm{\pi}$ complies with \eqref{eq:constraint_convergence}, we have $\nabla \! f(\mathbf{x}_{\bm{\pi}}) \in \ell_2$ and $\mathbf{u}_{\bm{\pi}} \in \ell_2$ for all $x_0\in \mathbb{R}^d$. Hence, $\mathbf{V}(\bm{\delta}^{x_0})\in \ell_2$ for every $x_0\in \mathbb{R}^d$ and every $f \in \mathcal{S}_\beta$ as the sum of two signals in $\ell_2$ lies in $\ell_2$.
    
    It remains to prove that \eqref{eq:V_CL} is equivalent to \eqref{eq:induced_by_pi} for $\mathbf{V}$ chosen as per \eqref{eq:choice_of_V}. We prove this by induction with a similar proof method as \cite[Theorem~2]{furieri2022neural}. For the closed-loop mapping \eqref{eq:induced_by_pi}, we define $\bm{\Psi}=(\Psix,\Psiu,\Psig)$ such that $\bm{\Psix}(f,\deltab) = \mathbf{x}_{\bm{\pi}}$, $\bm{\Psiu}(f,\deltab) = \mathbf{u}_{\bm{\pi}}$ and $\bm{\Psig}(f,\deltab) = \nabla \! f(\mathbf{x}_{\bm{\pi}})$. Similarly, for the closed-loop mapping \eqref{eq:V_CL} corresponding to $\mathbf{V}(\deltab)$ in \eqref{eq:choice_of_V}, we define $\bm{\Phi}=(\Phix,\Phiu,\Phig)$ such that $\bm{\Phix}(f,\deltab) = \mathbf{x}$, $\bm{\Phiu}(f,\deltab) = -\eta \nabla \! f(\mathbf{x})+\mathbf{V}(\deltab)$ and $\bm{\Phig}(f,\deltab) = \nabla \! f(\mathbf{x})$. For the inductive step, we assume that, for any $j \in \mathbb{N}$, we have $\Phi^u_i = \Psi^u_i$, $\Phi^x_i = \Psi^x_i$ and $\Phi^{\! \nabla}_i = \Psi^{\nabla}_i$ for all $i \in \mathbb{N}$ with $0\leq i \leq j$. Note that \eqref{eq:algorithm_dynamics} implies $\Phi^x_{j+1} = \Phi^x_{j} + \Phi^u_{j} + I$ and $\Psi^x_{j+1} = \Psi^x_{j} + \Psi^u_{j} + I$, 
    which ensure that $\Phi^x_{j+1} = \Psi^x_{j+1}$ by inductive assumption. Hence, it follows that $\Phi^{\! \nabla}_{j+1} = \nabla \! f(\Phi^x_{j+1}) = \nabla \! f(\Psi^x_{j+1}) = \Psi^{\! \nabla}_{j+1}$. For $\mathbf{V}$ chosen as per \eqref{eq:choice_of_V}, we also have $\Phi_{j+1}^u = -\eta \nabla \! f(\Phi^x_{j+1}) + \eta \nabla \! f(\Psi^x_{j+1}) + \Psi_{j+1}^u$,
    which simplifies to $\Phi^u_{j+1} = \Psi^u_{j+1}$. For the base case $j = 0$, we have $\Phi_0^x = \Psi_0^x = I$ by inspection of \eqref{eq:algorithm_dynamics_operator_form}, and thus $\Phi^{\! \nabla}_{0} = \nabla \! f(\Phi_0^x) = \nabla \! f(\Psi_0^x) = \Psi^{\! \nabla}_{0}$. Last, we also have $\Phi^u_0 = -\eta \nabla \! f(\Phi^x_{0}) + \eta \nabla \! f(\Psi^x_{0}) + \Psi_{0}^u = \Psi_{0}^u$.
\end{proof}

\begin{proof}[Theorem~\ref{th:reformulation}]
    If $0 < \eta < \beta^{-1}$, any algorithm in the form $\bm{\pi}(f,\mathbf{x}) = -\eta \nabla f(\mathbf{x})+\mathbf{v}$, with $\mathbf{v} \in \ell_2$, belongs to $\Sigma(f)$ for any $f \in \mathcal{S}_\beta$ by Lemma~\ref{le:sufficiency}. Since $\mathbf{V} \in \mathcal{L}_2$ and $\bm{\delta}^{x_0} \in \ell_2$, we have that $\mathbf{v} = \mathbf{V}(\bm{\delta}^{x_0}) \in \ell_2$, that is, any $\mathbf{V} \in \mathcal{L}_2$ leads to a feasible solution of \eqref{eq:metaopt_constrained_problem}. Hence, \eqref{eq:constraint_convergence} becomes redundant, and can thus be removed from \eqref{eq:metaopt_constrained_problem_L2_operator}, after plugging the update rule \eqref{eq:V_CL} in \eqref{eq:algo_def_metaloss_problem}.     Last, Lemma~\ref{le:necessity} ensures that for any $x_0 \in \mathbb{R}^d$, $f \in \mathcal{S}_\beta$, and $\bm{\pi}(f, \xb) \in \Sigma(f)$, there exists $\mathbf{V} \in \mathcal{L}_2$ such that the update rule in \eqref{eq:V_CL} reproduces the trajectories of $\bm{\pi}(f, \xb)$. In turn, any feasible solution of \eqref{eq:metaopt_constrained_problem} is considered in \eqref{eq:metaopt_constrained_problem_L2_operator}, implying equivalence.
\end{proof}

\begin{proof}[Theorem~\ref{th:sufficiency_batch}]
Rolling out \eqref{eq:algorithm_dynamics} for $M$ steps, starting from any $t$ such that $t \bmod M = 0$, and using the update rule \eqref{eq:update_batch} with $\tau = \left \lfloor t/M \right \rfloor$, we obtain the iteration:
\begin{equation}
    x_{t+M} = x_t - \eta_{\tau} \nabla \! f(x_t) + \eta_{\tau} w_t - \eta_{\tau} \sum\nolimits_{i = 0}^{M-1} v_{t+i}\,, \label{eq:M_step_forward_recursion}
\end{equation}
where the term $w_t$, which represents an error in the gradient direction relative to the gradient iteration \eqref{eq:separation}, is given by $w_t = \sum\nolimits_{i = 1}^{M-1}  \nabla \! f_{i}(x_t) - \nabla \! f_{i} (x_{t+i})$.
Under the assumptions of Lemma~\ref{th:sufficiency_batch}, we have that
\begin{equation}
    \label{eq:batch_M_vt_bound}
    \left|\sum\nolimits_{i = 0}^{M\!-\!1} \! \! v_{t+i}\right| \!
    \leq \! \sum\nolimits_{i = 0}^{M\!-\!1} \! |v_{t+i}| \! \leq \! \eta_{\tau} M (C\!+\!s D |\nabla \! f (x_t)|)\!\,;
\end{equation}
we now proceed to show that a similar bound also holds for $|w_t|$. %
By the triangle inequality and observing that $f_i \in \mathcal{S}_\beta$ ensures that $\nabla \! f_i$ is $\beta$-Lipschitz continuous, we have that $|w_t| \leq \! \sum_{i = 1}^{M-1} \! |\nabla \! f_i (x_t) - \nabla \! f_i (x_{t+i})| \leq \beta \! \sum_{i = 1}^{M-1} \! |x_t - x_{t+i}|$, which further implies:
\begin{align}
    |w_t| \leq \eta_{\tau} \beta |\nabla \! f_0(x_t) + v_t| + \beta \sum\nolimits_{i = 2}^{M-1} |x_t - x_{t+i}|\,,\label{eq:norm_bound_error_gradient_direction_intermediate}
\end{align}
where the term $\sum_{i = 2}^{M-1} |x_t - x_{t+i}|$ is given by:
\begin{equation}
    \label{eq:norm_bound_error_gradient_direction_double_sum}
    \hspace{-0.3cm}\eta_{\tau} \sum\nolimits_{i = 2}^{M\!-\!1} \! \left|\nabla \! f_0(x_t) + v_t s+ \sum\nolimits_{j = 1}^{i\!-\!1} \nabla \! f_{j}(x_{t+j}) \!+\! v_{t+j}\right| \,.
\end{equation}
Then, we observe that $\eqref{eq:norm_bound_error_gradient_direction_double_sum}$ can be upper-bounded by $\eta_{\tau} (M-2)|\nabla \! f_0(x_t) + v_t| + \eta_{\tau} \sum_{i = 2}^{M-1} \sum_{j = 1}^{i-1} |\nabla \! f_{j}(x_{t+j}) + v_{t+j}| \leq \! \eta_{\tau} (M - 2) \! \left( \! |\nabla \! f_0(x_t) + v_t| + \! \! \sum_{i = 1}^{M-2} \! |\nabla \! f_{i}(x_{t+i}) + v_{t+i}| \! \right)$.
By combining these inequality with \eqref{eq:norm_bound_error_gradient_direction_intermediate}, we deduce that $\eta_{\tau} \beta (M\!-\!1) \! \left(\! |\nabla \! f_0(x_t)| \!+\! \! \sum\nolimits_{i = 0}^{M\!-\!2} \! |v_{t+i}| \! + \! \sum\nolimits_{i = 1}^{M\!-\!2} \! |\nabla \! f_{i}(x_{t+i})| \! \right)$ upper-bounds $|w_t|$.
Moreover, for any $i = 1, \ldots, M-2$, it holds that $|\nabla \! f_i(x_{t+i})| \!=\! |\nabla \! f_i(x_{t+i-1} - \eta_{\tau} (\nabla \! f_{i-1}(x_{t+i-1}) + v_{t+i-1}))| \leq |\nabla \! f_i(x_{t+i-1})| + \eta_{\tau} \beta |\nabla \! f_{i-1}(x_{t+i-1})| + \eta_{\tau} \beta |v_{t+i-1}|$.
By iterating the reasoning above, we have that $ |\nabla \! f_i(x_{t+i})|$ is upper-bounded by $\sum_{j = 0}^i E |\nabla \! f_j(x_{t})| + F |v_{t+j}|$ for appropriately defined positive constant $E$ and $F$.
Finally, leveraging \eqref{eq:assumption_batch_gradients} and \eqref{eq:batch_M_vt_bound}, we conclude that there exists positive constant $P$ and $Q$ such that $\left|w_t - \sum_{i = 0}^{M-1} v_{t+i}\right| \leq \eta_{\tau} (P + Q |\nabla \! f(x_t)|)$.
Having established this upper bound, the results of Lemma~\ref{th:sufficiency_batch} follow from applying \cite[Proposition~1]{bertsekas2000gradient} to the recursion \eqref{eq:M_step_forward_recursion}, since $f$ is bounded from below.
\end{proof}

\end{document}